\documentclass[times,aps,pre,twocolumn,groupedaddress,showpacs]{revtex4-1}  
\newcommand{\width}{8cm}

\usepackage{graphicx} 
\usepackage{dcolumn} 
\usepackage{bm} 
\usepackage{times}
\usepackage{colordvi}

\begin{document}

\newcommand{\FIGURE}[2]{
       \typeout{FIGURE #1.eps}
       \begin{figure}[h]
       \includegraphics[width=\width]{#1.eps}
       \caption{#2}
       \label{#1}
       \end{figure}
}

\newcommand{\Fbox}[1]{\mbox{$\bigcirc\!\!\!\!${\scriptsize #1}$\;$}}


\title{Finite size scaling in asymmetric systems of percolating sticks}

\newcommand{\SCL}{Scientific Computing Laboratory, Institute of Physics Belgrade,
University of Belgrade, 11080 Belgrade, Serbia}

\author{Milan \v{Z}e\v{z}elj}
\email[Corresponding author: ]{milan.zezelj@ipb.ac.rs}
\affiliation{\SCL}
\author{Igor Stankovi\'{c}}
\affiliation{\SCL}
\author{Aleksandar Beli\'{c}}
\affiliation{\SCL}

\begin{abstract}
We investigate finite size scaling in percolating widthless stick
systems with variable aspect ratios in an extensive Monte Carlo
simulation study. A generalized scaling function is introduced to
describe the scaling behavior of the percolation distribution
moments and probability at the percolation threshold. We show that
the prefactors in the generalized scaling function depend on the
system aspect ratio and exhibit features that are generic to whole
class of the percolating systems. In particular, we demonstrate
existence of characteristic aspect ratio for which percolation
probability at the threshold is scale invariant and definite
parity of the prefactors in generalized scaling function for the
first two percolation probability moments.
\end{abstract}

\keywords{percolation, nanorod systems, scaling phenomena, transport properties}

\pacs{89.75.Da, 72.80.Tm, 05.10.Ln}


\maketitle

\section{INTRODUCTION}
Recently there is an increasing interest in the randomly
distributed stick (rodlike)
particles~\cite{2003_apl_Ramasubramaniam, 2005_apl_Bo,
2009_prl_Trionfi, 2010_prb_Hazama, 2010_apl_Sangwan}, due to
promising developments in the area of the conducting rodlike
nanoparticle networks, such as carbon nanotubes, silicon, copper,
and silver nanowires, with applications in
electronics~\cite{2010_apl_Sangwan, 2008_apl_Li, 2005_prl_Kumar},
optoelectronics~\cite{2004_nl_Hu}, and
sensors~\cite{2008_nature_Cao, 2009_prb_Li}. On the theoretical
side, most of the work done until now in the field of percolation
of random networks is done for lattice percolation~\cite{1981_pla_Stauffer,
1992_jpa_Cardy, 1992_prl_Ziff, 1994_prl_Aharony, 1996_pre_Hovi,
2000_prl_Newman, 2002_pre_Ziff, 2003_ipt_Stauffer}. The random stick networks
are an important representative of continuum percolation~\cite{1974_prb_Pike,
1983_prb_Balberg, 1983_prl_Balberg, 1984_prb_Balberg}. Random
stick percolation and lattice percolation fall into the same
universality class having the same critical exponents~\cite{1983_prl_Balberg}.
Previous studies established that all
systems fall on the same scaling function if dimensionality of the
system, percolation rule, boundary conditions, and aspect ratio
are fixed~\cite{1996_pre_Hovi}. In applications, aspect ratio of
the rectangular system is the only variable parameter, e.g.,
geometry of transistor gate channel in the carbon nanotube
transistors~\cite{2005_apl_Bo, 2010_apl_Sangwan}. The objective of
the present paper is to describe in consistent way finite size
scaling of average percolation density and standard deviation for the
asymmetric rectangular stick systems with free boundaries. From
general scaling arguments one would expect that for all finite
size systems their convergence is governed by an exponent
$-1/\nu$~\cite{2003_ipt_Stauffer}. For two-dimensional (2D)
systems $\nu=4/3$~\cite{2003_ipt_Stauffer}. Following initial
Ziff's publication~\cite{1992_prl_Ziff}, Hovi and
Aharony~\cite{1994_prl_Aharony, 1996_pre_Hovi} argued that the
irrelevant scaling variables in the renormalization-group
treatment of percolation imply a slower leading-order convergence
of percolation  probability to its infinite-system value,
characterized by an exponent $-1/\nu-\theta$, whose value was
deducted from the Monte Carlo work of Stauffer to be $\theta
\approx 0.85$~\cite{1981_pla_Stauffer}. Further it was showed that
for lattice percolation on the square system leading exponent of
the average concentration at which percolation first occurs is
$-1/\nu-\theta$, where $\theta \approx 0.9$~\cite{2002_pre_Ziff}.
All the previous studies were performed for symmetric systems. We
show that only in symmetric case the exponent of average density
is $-1/\nu-\theta$. In asymmetric systems, we observe a leading
$-1/\nu$ exponent. Another quantity, the percolation probability
at the percolation threshold in symmetric bond percolating systems is size
independent, i.e., scale invariant~\cite{1978_prb_Bernasconi}.
Until now, this behavior is not observed in other types of random
percolating systems. We will demonstrate that asymmetric systems
can exhibit scale invariant behavior.

In this paper, we investigate finite size scaling of the
asymmetric rectangular stick systems with free boundaries. Both 
from renormalization group considerations and in the simulations, 
we find that aspect ratio strongly influences scaling behavior of 
the percolation probability distribution function moments, i.e.,
average density of sticks at which percolation first occurs and
variance of percolation probability distribution function. A
generalized scaling function is introduced, with aspect ratio
dependent prefactors and constant exponents of the expansion.
Finally, it is shown that percolation probability of the
asymmetric infinite sticks system at the critical threshold
density agrees with Cardy's analytic formula~\cite{1992_jpa_Cardy}.

\section{NUMERICAL METHOD FOR CALCULATION OF PERCOLATION PROBABILITY}
Monte Carlo simulations, coupled with an efficient cluster
analysis algorithm and implemented on grid platform, are used to
investigate the stick percolation~\cite{2001_pre_Newman,
2002_cpc_Stankovic, 2011_jgc_Balaz, 2009_pre_Li}. We consider two
dimensional (2D) systems with isotropically placed widthless
sticks. The sticks of unit length are randomly positioned and
oriented inside the rectangular system of width $L_x$ and height
$L_y$. Two sticks lie in the same cluster if they intersect.
System percolates if two opposite boundaries are connected with
the same cluster. The aspect ratio $r$ is defined as the length of
the rectangular system in percolating direction divided with the
length in perpendicular direction. We define the normalized system
size as a square root of the rectangular area $L=\sqrt{L_xL_y}$
(geometric average), which represent the length of the square
system with the same area. The behavior of stick percolation is
studied in terms of the number stick density $n=N/L^2$. The
percolation threshold of infinite system is defined by the
critical density $n_c \approx 5.63726$~\cite{2009_pre_Li}. Monte
Carlo simulations are performed for wide range of the aspect
ratios, $0.1 \leq r \leq 10$. In order to ensure the
same precision for small and large systems we collected more than
$N_{MC}=10^9$ Monte Carlo realizations for small systems $L < 10$,
down to $N_{MC}=10^7$ for the largest system $L = 320$. Using
appropriate functions for fitting data and the least squares
fitting methods excellent fits were obtained ($R^2 > 0.9999$) for
all analyzed systems with $L\geq 16$. Statistical errors for the
calculations are estimated in conventional fashion using standard
deviation~\cite{2001_pre_Newman}.

Percolation probability function $R_{N,L,r}$, is the probability that
the system with $N$ sticks, size $L$ and aspect ratio $r$ percolates.
It is convenient to pass from the discrete percolation probability
function $R_{N,L,r}$ for $N$ sticks to a probability function for
arbitrary stick density $n$~\cite{2009_pre_Li}
\begin{equation}
R_{n,L,r}=\sum_{N=0}^\infty\frac{(nL^2)^Ne^{-nL^2}}{N!}R_{N,L,r}, \label{RnLr}
\end{equation}
with percolation probability distribution function defined as
$P_{n,L,r} = \partial R_{n,L,r} / \partial n$. The average stick
percolation density at which, for the first time, a percolating
cluster connects boundaries of the system is
\begin{equation}
\langle n \rangle_{L,r}=\int_0^\infty \!\! {nP_{n,L,r}dn}=
\frac{1}{L^2}\sum_{N=0}^\infty(1-R_{N,L,r}), \label{n_av}
\end{equation}
where the last equality follows from integrating by parts. Another
important parameter of probability distribution function,
$P_{n,L,r}$, is variance $\Delta^2_{L,r}=\langle n^2
\rangle_{L,r}-\langle n \rangle^2_{L,r}$, where $\langle n^2
\rangle_{L,r}$ is calculated as
\begin{equation}
\langle n^2 \rangle_{L,r}\!=\!\!\int_0^\infty \!\!\! {n^2 P_{n,L,r}dn}\!=\!
\frac{2}{L^4}\!\sum_{N=0}^\infty\!(N\!+\!1)(1\!-\!R_{N,L,r}). \label{n2_av}
\end{equation}
Equations~(\ref{n_av}) and~(\ref{n2_av}) allow calculations of the
first two moments directly from discrete percolation probability function
$R_{N,L,r}$. This is computationally more efficient since it
avoids calculation of function $R_{n,L,r}$ with high resolution.

\section{GENERALIZED SCALING FUNCTIONS FOR MOMENTS}
The percolation probability function is related to the universal
scaling function~\cite{1996_pre_Hovi}
\begin{equation}
R_{n, L, r} = F(\hat{x}, \{\hat{y}_i\}, \hat{z}).
\end{equation}
The arguments of the universal scaling function $F$ are $\hat{x}=
A (n - n_c)L^{1/\nu}$, $\hat{y}_i = B_i\omega_i L^{-\theta_i}$,
and $\hat{z} = C\ln(r)$, where $A$, $\{B_i\}$, and $C$ are the
nonuniversal metric factors, $\{\omega_i\}$ are the irrelevant
variables, and $\{\theta_i\}$ are the corrections to scaling
exponents, ($i = 1,2,...)$. Using free boundary conditions and
considering two complementary systems: the sticks and empty space
around the sticks, we can conclude that either the sticks
percolate in one direction or the empty space percolates in the
opposite direction:
\begin{equation}
F(\hat{x}, \{\hat{y}_i\}, \hat{z}) + F(-\hat{x}, \{-\hat{y}_i\}, -\hat{z}) = 1.
\end{equation}
Taking the derivative with respect to $\hat{x}$, $\hat{y}_i$, and
$\hat{z}$ and evaluating the derivatives at
$\hat{x}=\hat{y}_i=\hat{z}=0$, (i.e., at $n=n_c,L\to\infty,r=1$),
we conclude that $\partial^m
F/\partial\hat{x}^{j}\partial\hat{y}_1^{k_1}
...\left.\partial\hat{z}^l\right|_0=0$, for $m$ even. Expanding
the percolation probability function near the critical point we
find that
\begin{equation}
F(\hat{x}, \{\hat{y}_i\}, \hat{z})\!=\!F(0, \{0\}, 0)\!+\!f_0(\hat{x}, \hat{z})\!+\!\sum_{i=1}^{\infty}
f_i(\hat{x},\hat{z})\hat{y}_i\!+\!.... \label{F_series}
\end{equation}
where the functions $f_0(\hat{x}, \hat{z})$ and $f_i(\hat{x},
\hat{z})$ are defined by
\begin{equation}
f_0(\hat{x}, \hat{z}) =
\sum_{j,l=0}^{\infty}\frac{1}{j!l!}\left.\frac{\partial^{j+l}F}{\partial\hat{x}^{j}\partial\hat{z}^l}
\right|_0\hat{x}^{j}\hat{z}^l, \text{ for } j+l \text{ odd}, \label{f_0}
\end{equation}
and
\begin{equation}
f_i(\hat{x}, \hat{z}) = \sum_{j,l=0}^{\infty}\frac{1}{j!l!}
\left.\frac{\partial^{j+l+1}F}{\partial\hat{x}^{j}\partial\hat{y_i}\partial\hat{z}^l}\right|_0\hat{x}^{j}\hat{z}
^l, \text{ for } j+l \text{ even}. \label{f_1}
\end{equation}
Since percolation probability distribution function $P_{n,L,r} =
\partial R_{n,L,r} / \partial n$ gives the probability
distribution for a system of size $L$ and aspect ratio $r$ to
percolate for the first time at stick density $n$, we can define
the moments of this distribution
\begin{eqnarray}
\mu_k &=& \int_0^{\infty} (n - n_c)^k \frac{\partial R_{n, L, r}}{\partial n} dn \nonumber \\ &=&
A^{-k}L^{-k/\nu}\int_{-An_cL^{1/\nu}}^{\infty} \hat{x}^k\frac{\partial F}{\partial \hat{x}} d\hat{x},
\label{mu_fin}
\end{eqnarray}
Substituting Eqs.~(\ref{F_series})-(\ref{f_1}) in Eq.~(\ref{mu_fin}) the $k$th moment scales as
\begin{equation}
\mu_k(\{\hat{y}_i\}, \hat{z}) = L^{-k/\nu} \left( g_0(\hat{z}) + \sum_{i=1}^{\infty} g_i(\hat{z}) \hat{y}_i + ...
\right), \label{mu_k}
\end{equation}
where we introduce general functions $g$. For odd $k$,
$g_0(\hat{z})$ is odd function and $g_i(\hat{z})$ are even
functions of $\hat{z}$. For even $k$, $g_0(\hat{z})$ is even and
$g_i(\hat{z})$ are odd functions. Therefore, observed parity of 
prefactors in respect to $\hat{z}$ should be independent of the type of
the system.

From Eq.~(\ref{mu_k}) the scaling behavior of the $\langle n
\rangle_{L,r}$ can be described with generalized moment scaling
function with aspect ratio dependent coefficients
\begin{equation}
\langle n \rangle_{L,r} = n_c + L^{-1/\nu}\sum_{i=0}^\infty a_i(r) L^{-\theta_i}.
\label{n_av_series}
\end{equation}
where $\{\theta_i\}$ are the corrections to scaling exponents. The
zeroth-order correction to exponent $\theta_0$ should be zero~\cite{2003_ipt_Stauffer}.
In analogy to $\langle n \rangle_{L,r}$, for variance $\Delta^2_{L,r}$ we introduce
expansion
\begin{equation}
\Delta^2_{L,r} = L^{-2/\nu}\sum_{i=0}^\infty b_i(r) L^{-\theta_i}. \label{Delta2_series}
\end{equation}
From Eq.~\ref{mu_k} and the parity of $g_0(\hat{z})$ and
$g_i(\hat{z})$, for the zeroth-order and the first-order prefactors
for $\langle n \rangle_{L,r}$ and $\Delta^2_{L,r}$ near $\ln(r) =
0$ (i.e., $\hat{z}=0$), we obtain approximate expressions for
$a_0(r)\approx a_{0,0}\ln(r)+a_{0, 1}\ln^3(r)$, $a_1(r)\approx
a_{1, 0}+a_{1, 1}\ln^2(r)$, $b_0(r) \approx b_{0, 0} + b_{1,
0}\ln^2(r)$, $b_1(r) \approx b_{1, 0}\ln(r) + b_{1, 1}\ln^3(r)$.

\section{RESULTS AND DISCUSSION}
The results for percolation probability $R_{n,L,r}$ and
distribution $P_{n,L,r}$ function are shown in
Fig.~\ref{figure_1}. One observes that the slope of percolation
probability function increases with the increase of the system
size. The percolation probability function curves intersect
approximately at $n_c$. The fine behavior of percolation
probability at $n_c$ will be discussed bellow. With the increasing
system size, the standard deviation of probability distribution function
decreases to zero. Also, average stick percolation density
$\langle n \rangle_{L,r}$, which corresponds roughly to maximum of
probability distribution function $P_{n,L,r}$, approaches to
percolation threshold $n_c$. For $r<1$, $\langle n \rangle_{L,r}$
converges to $n_c$ from below with increase of the system size
$L$. The reason for this is that narrow finite systems will be
spanned already at lower densities than $n_c$. For $r>1$, $\langle
n \rangle_{L,r}$ converges from above, while for symmetric systems
($r=1$) is roughly centered at $n_c$, see Fig~\ref{figure_1}. From
Fig.~\ref{figure_2}, one can see that average stick percolation
density $\langle n \rangle_{L,r}$ for aspect ratio higher than one
is monotonically decreasing function of the system size $L$.
Somewhat surprising, for aspect ratios lower than one, $\langle n
\rangle_{L,r}$ is not monotonic function and has local minimum, i.e.,
for small systems $\langle n \rangle_{L,r}$ is a decreasing
function, which passes through $n_c$, reaches a minimum and after
that converges to $n_c$ from below. In inset of
Fig.~\ref{figure_2}, one can see that for large system sizes all
the curves show power law convergence to the percolation threshold
$n_c$ with exponent $-1/\nu$, except in the symmetric case, i.e.,
$r=1$, where exponent is $-1/\nu-\theta_1$. Absolute values of the
leading-order prefactors are the same for aspect ratios $r$ and
$1/r$. Higher exponent of symmetric systems comes from the basic 
physics of percolation that is connectedness. We can illustrate this 
using a simplified image of site percolation by introducing quantity 
$R(p)$ as probability that the sites with occupancy $p$ form a spanning 
path. The percolation probability $R(p)$ and occupancy $p$ are equivalent to 
the percolation probability function $R_{n,L,r}$ and stick density $n$, 
respectively. In this image, a cell coming out of the renormalization 
(coarse graining) transformation is occupied only if it 
contains a set of sites that span this cell. The universal scaling 
function in previous section reflects the fact that the probability of 
the spanning system at the percolation threshold $R(p_c)$ remains unaltered 
under this transformation~\cite{2003_ipt_Stauffer}. Therefore fixed point of this system, i.e., 
the critical percolation threshold, $p_c$ is satisfying relation 
$p_c= R(p_c)$. We can expand the percolation probability around the 
percolation threshold $p_c$, $|R(p)-R(p_c)|\approx dR/dp|_{p_c}|p-p_c|$. Also, 
if we renormalize lattice by a length factor $b$, close to $p_c$, 
characteristic length changes as $\xi/b$. Since $\xi\sim|p-p_c|^{-\gamma}$, 
we can write another relation, $|R(p)-R(p_c)|^{-\gamma} \approx
|p-p_c|^{-\gamma}/b$, connecting characteristic lengths before and
after renormalization. From these two relations one can conclude
that critical exponent should be
\begin{equation}
-1/\gamma\approx \frac{\ln dR/dp|_{p_c}}{\ln 1/b}.\label{loglog}
\end{equation}
From Fig~\ref{figure_1}, one can see that probability
density $P_{n_c,L,r}$ which is derivative of $R_{n,L,r}$ at $n_c$,
is always larger for symmetric systems than for asymmetric systems
of the same size. Therefore, from Eq.~(\ref{loglog}), one expects
higher absolute value of the exponent in symmetric compared to
asymmetric systems. Another conclusion one can draw from this
analysis, is that the observed exponents are a result of the
interplay of the characteristic length and the system shape. Usually, 
such behavior is attributed to a competition between two-dimensional
and three-dimensional (or one-dimensional and two dimensional), e.g., 
in Ising model for slab geometries, cf. Ref~\cite{2005_plischke}. In 
this system we observe that there is sharp transition in nature of 
scaling when we pass from symmetric to asymmetric system, and a 
competition between exponents characteristic for symmetric and asymmetric 
systems.

\FIGURE{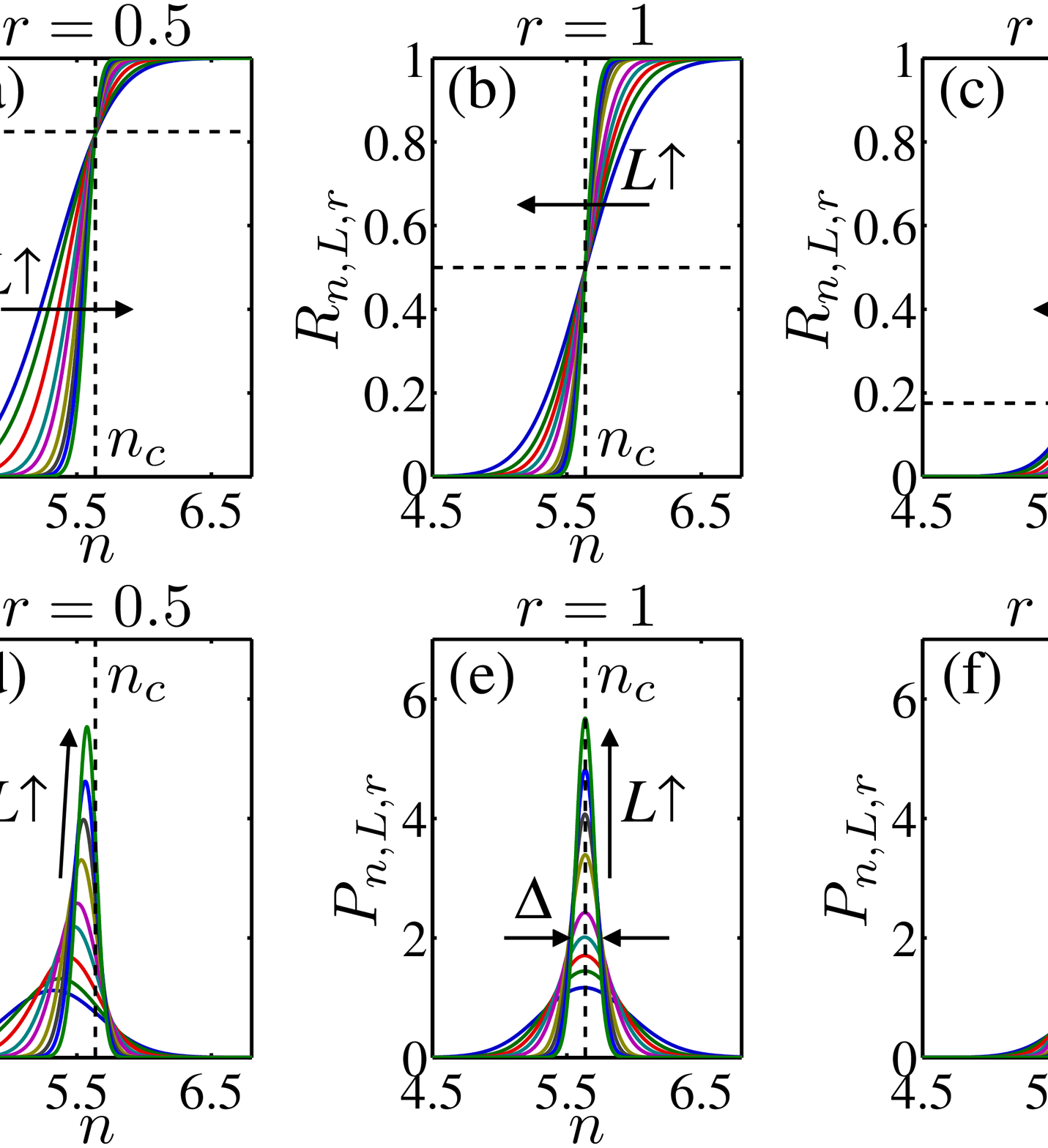}{(Color online) Percolation probability function $R_{n,L,r}$
(a, b, c) and probability distribution function $P_{n,L,r}$ (d, e, f)
for stick percolation on rectangular systems with free boundary conditions
and increasing system size from $L=20$ to $200$ for three aspect
ratios $r = 0.5, 1$, and $2$. The direction of the increase of $L$ is 
indicated on graphs. The vertical dashed lines denote the value for the
percolation threshold $n_c$, while the horizontal dashed lines (a, b, c)
denote the percolation probability on the infinite systems at
the threshold $R_{n_c, L\rightarrow\infty,r}$.}

\FIGURE{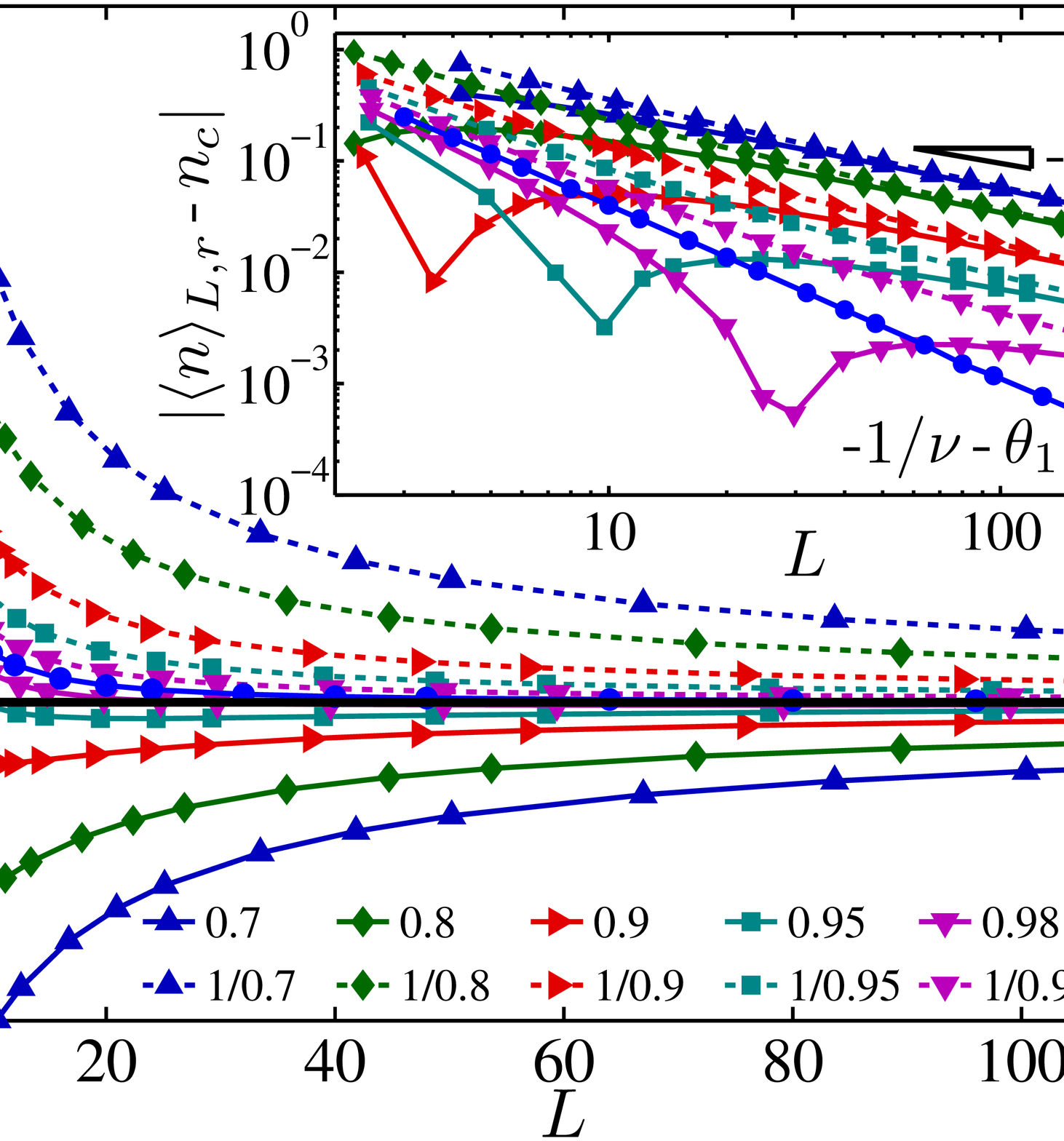}{(Color online) The dependence of the average stick
percolation density $\langle n \rangle_{L,r}$ on the system size $L$
and aspect ratio $r$. The values are obtained from Monte Carlo simulations and
calculated using Eq.~(\ref{n_av}). The values are given for aspect
ratios $r = 0.7, 0.8, 0.9, 0.95, 0.98, 1$ (solid lines) and their
inverse values $r = 1/ 0.7, 1/ 0.8, 1/ 0.9, 1/ 0.95, 1/ 0.98$
(dashed lines). The horizontal bold line denotes the expected value for the
percolation threshold $n_c$. Inset: The same data is shown in
logarithmic scale to demonstrate the same power law convergence of
the $r$ and $1/r$ pairs.}

From Monte Carlo simulation data we have obtained the two
leading-order terms of $\langle n \rangle_{L,r}$ in
Eq.~(\ref{n_av_series}) by interpolation, cf. Ref.~\cite{my_1}.
Results of the analysis are shown in Fig.~\ref{figure_3}. The
zeroth-order prefactor is zero for symmetric system $r=1$, and it
is odd function on a logarithmic scale, i.e., $a_0(r) = -a_0(1/r)$.
We have verified the obtained results by interpolation through
symmetrizing data points $(\langle n \rangle_{L,r}+\langle
n\rangle_{L,1/r})/2$. The fitting coefficients for prefactors
$a_{i,j}$ are calculated using the least squares fitting methods
and given in Table~\ref{my_table}, cf. Fig.~\ref{figure_3}.
Influence of higher order terms were comparable or smaller than
simulation data error and we could not extract them with
sufficient precision. For the first-order correction, we obtain
$\theta_1=0.83(2)$ for $r = 1$, cf. Ref.~\cite{1981_pla_Stauffer,
1992_prl_Ziff}. The residual aspect ratio dependence of $\theta_1$
cannot be further analyzed without provision of retaining the
first two terms in Eq.~(\ref{n_av_series}). The system size where
the average density reaches minimum is $L_{\min}\approx
\left(-a_1(r)/a_0(r)(1+\nu\theta_1)\right)^{1/\theta_1}$, cf.
Eq.~(\ref{n_av_series}). For narrow systems, $r<1$, $L_{\min}$
diverges as $1/\ln(r)$ as $r$ approaches one. For $L<L_{\rm min}$
the first-order term is dominant.

\FIGURE{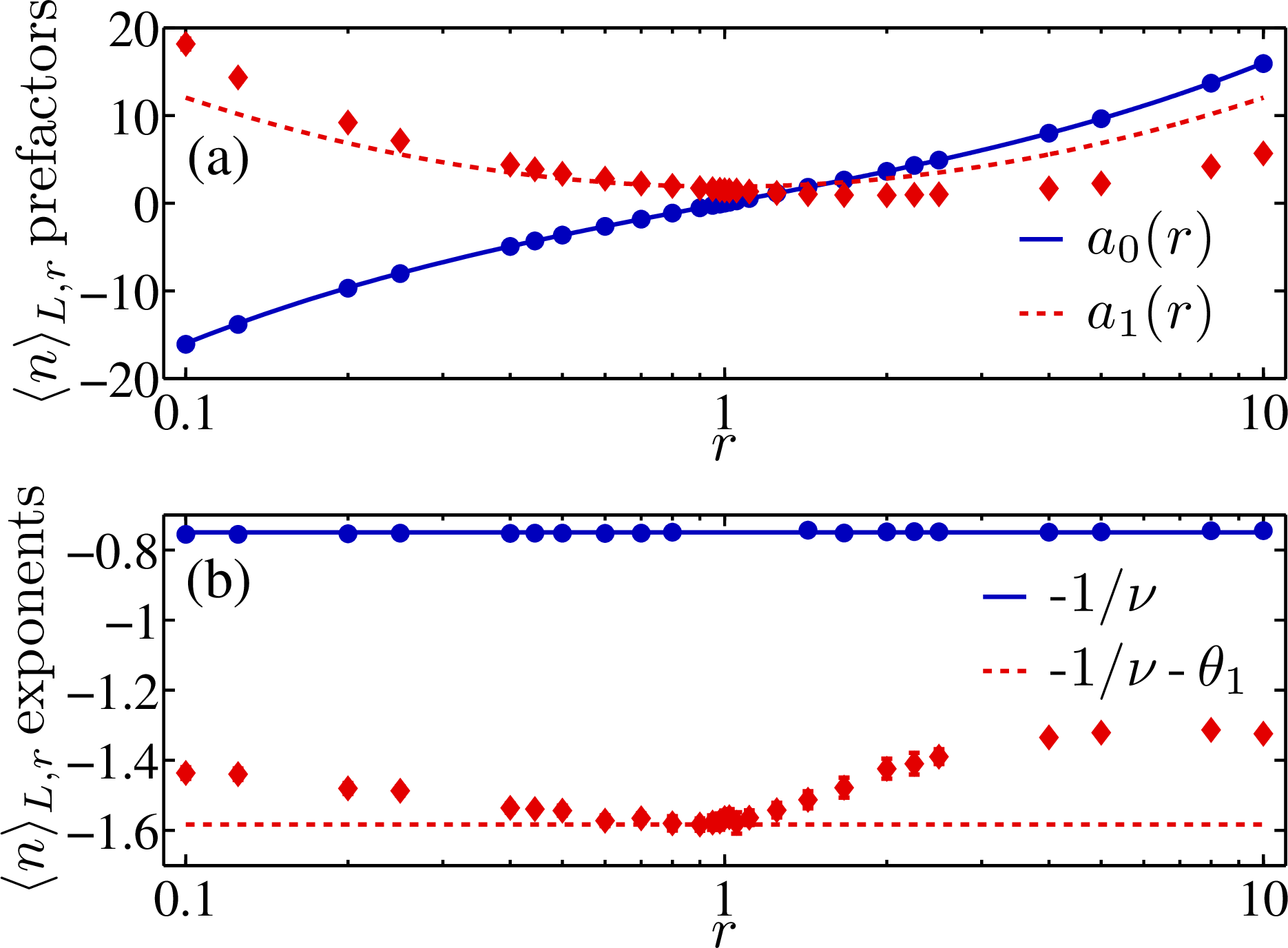}{(Color online) Prefactors (a) and exponents (b) are
shown for the two leading-order terms of generalized scaling function
for average stick percolation density $\langle n \rangle_{L,r}$,
Eq.~(\ref{n_av_series}). The zeroth-order prefactor is odd function on a
logarithmic scale, i.e., $a_0(r) = -a_0(1/r)$, and the zeroth-order exponent
is $-1/\nu$ (solid lines). The first-order prefactor is even function,
i.e., $a_1(r) = a_1(1/r)$ and the first-order correction to the scaling
exponent is $\theta_1=0.83(2)$ for $r = 1$ (dashed lines).}

\begin{center}
\begin{table}[h]
\caption{Results for the coefficients $a_{i,j}$ and $b_{i, j}$,
where $i,j \in \{0, 1\}$. The results are obtained using the least
squares method.}
\begin{tabular}{|c|c|c|c|c|}
\hline
& 0, 0 & 0, 1 & 1, 0 & 1, 1 \\
\hline
$a_{i,j}$ & $5.08(1)$ & $0.352(4)$ & $1.9(5)$ & $1.9(6)$ \\
\hline
$b_{i,j}$ & $14.56(5)$ & $2.25(6)$ & $11(2)$ & $3(1)$ \\
\hline
\end{tabular}
\label{my_table}
\end{table}
\end{center}

The variance prefactors and exponents for the two leading-order
terms are shown in Fig.~\ref{figure_4}. The prefactors and
exponents are obtained by fitting $\Delta^2_{L,r}$ with the first
two terms in Eq.~(\ref{Delta2_series}), cf. Ref.~\cite{my_2}. The
fitting coefficients $b_{i,j}$ are given in Table~\ref{my_table}
and obtained prefactor dependences on $r$ are given in
Fig.~\ref{figure_4}. The zeroth-order prefactor of $\Delta^2_{L,r}$ 
is even function on a logarithmic scale, i.e., $b_0(r) = b_0(1/r)$,
as one can see from a coarse observation of the percolation
probability distribution function in Fig.~\ref{figure_1}. Asymmetry
of the variance, i.e. $\Delta^2_{L,r} \neq \Delta^2_{L,1/r}$, is the
first-order effect, cf. Eq.~(\ref{Delta2_series}).

\FIGURE{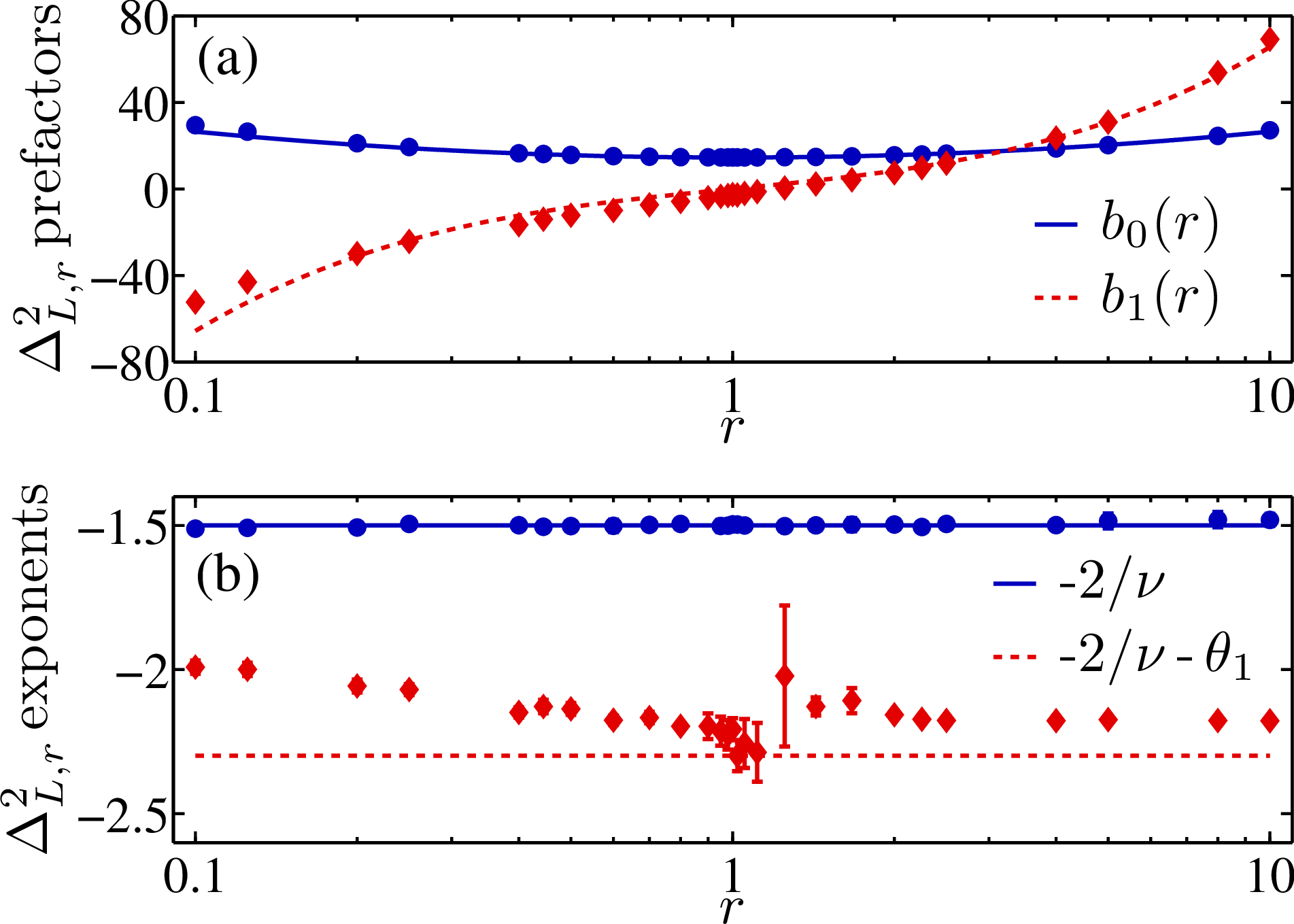}{(Color online) Prefactors (a) and exponents (b)
are shown for the two leading-order terms of generalized
scaling function for variance $\Delta^2_{L,r}$, Eq.~(\ref{Delta2_series}).
The zeroth-order prefactor is even function on a logarithmic scale, i.e.,
$b_0(r) = b_0(1/r)$ and the zeroth-order exponent is $-1/\nu$ (solid lines).
The first-order prefactor is odd function, i.e., $b_1(r) = -b_1(1/r)$,
and the first-order correction to the scaling exponent is $\theta_1=0.80(5)$
for $r = 1$ (dashed lines). Prefactor $b_1(r)$ passes through zero for
$r$ between 1/0.9 and 1/0.8 causing higher error bars of $\theta_1$.}

Finally, we investigate scaling behavior of percolation
probability at the percolation threshold ($R_{n_c, L, r}$), using
the generalized scaling function $R_{n_c, L, r}=R_{n_c,
L\rightarrow\infty, r}+c_1(r)/L+c_2(r)/L^2$, cf.
Ref.~\cite{2002_pre_Ziff}. The results for prefactors $c_{1}(r)$
and $c_{2}(r)$ are shown in Fig.~\ref{figure_5}. For two limiting
cases ($r<0.1$ and $r>10$), the prefactors are close to zero, what
is consistent with behavior observed in Fig.~\ref{figure_1}.
Between these two limiting cases, one can observe that both
prefactors are close to zero for $r=2.25(5)$. Furthermore, at this
aspect ratio, we could not observe existence of the higher order
terms. This means that at the percolation threshold percolation
probability is the same for all the systems, independent of system
size $L$, having the value $R_{n_c, r\approx 2.25}\approx 0.135$.
Scale invariance, i.e., $R_{n_c, L, r}=R_{n_c, \infty, r}$ is
already seen and intuitively understood for bond percolation in
symmetric systems, where $R_{p_c=0.5, r=1}=1/2$ is independent of
system size $L$, cf. Ref.~\cite{1978_prb_Bernasconi}. The reason
for observed system size invariance of percolation probability at
the threshold in the asymmetric stick system is existence of 
multiple zero of at least second order at this point in universal 
scaling function. Regarding value of percolation probability at 
the percolation threshold of infinite system, we find that Cardy's 
analytical model derived for the lattice percolation describes 
simulation data, cf. Fig.~\ref{figure_5}. The deviation between 
analytical value for lattice and Monte Carlo value for stick 
percolation is less than statistical error of the simulation data, 
i.e., less than $10^{-5}$. Percolation probability for 2D sticks 
system therefore satisfies $R_{n_c,L\rightarrow\infty, r}+R_{n_c,
L\rightarrow\infty, 1/r}=1$.

\FIGURE{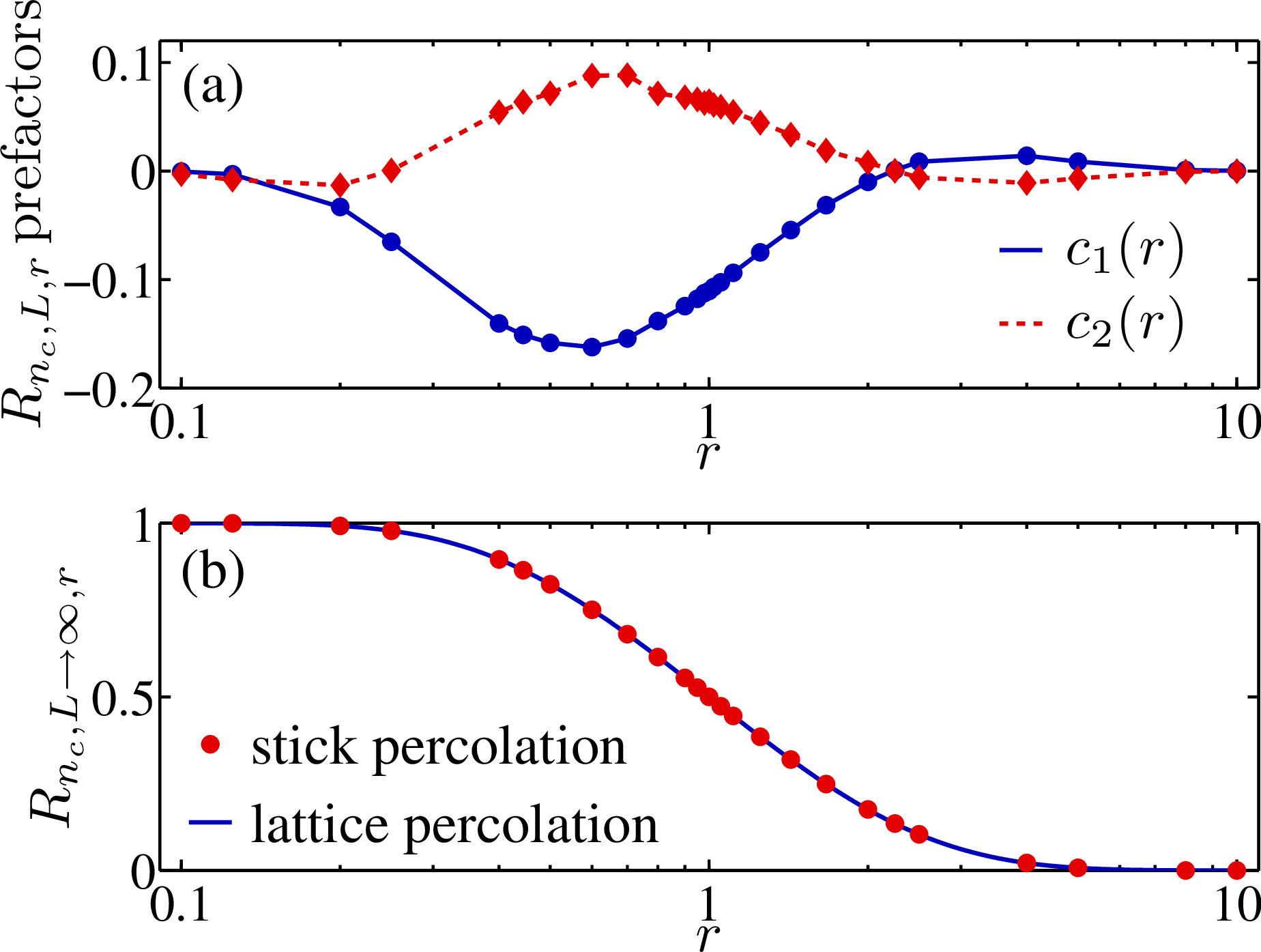}{(Color online) (a) Prefactors for finite size scaling of
the percolation probability at the percolation threshold are shown for
the two leading-order terms. (b) Percolation probability at the threshold
for infinitely large system $R_{n_c, L\rightarrow\infty, r}$. Points represent
Monte Carlo data for stick percolation, while line represents results of
Cardy's model for the lattice percolation. The error bars are much smaller
than the size of the symbols for all $r$}

\section{CONCLUSION}
In summary, based on the analysis of finite size scaling in
continuum two-dimensional systems, the generalized scaling law is
introduced for average percolation density, variance, and
percolation probability at the percolation threshold. The
presented methodology could be used to model accurately these
properties for any percolating system. We find that the zeroth-order 
prefactor of average percolation density is odd function with respect 
to $\ln(r)$. This explains the faster convergence of average
percolation density for symmetric systems than expected from
general scaling arguments. We also observe that there is a
characteristic aspect ratio for which percolation probability at
the percolation threshold is system size independent. In addition,
for infinite system, we find that percolation probability at the
critical threshold density shows excellent agreement with the
Cardy's prediction for lattice percolation. The presented results
confirm that continuum percolation belongs to the same
universality class as lattice percolation in the sense that value
of percolation probability at the threshold for infinitely large
systems is the same for lattice and continuum percolation. One
should note that a number of other features observed in this work
should be a common characteristic within the class, e.g.,
existence of the aspect ratio where the percolation probability at
the threshold is scale invariant and parity of the moments of the
percolation probability distribution function. This opens up the 
question of the particle shape influence on prefactors, weather it is 
possible to find systems where the observed behaviors are more 
pronounced, and finally the question of the general form of the 
prefactors for describing different systems.\\

\begin{acknowledgments}
The authors acknowledge support by the Ministry of Science of the
Republic of Serbia, under project No. ON171017. Numerical
simulations were run on the AEGIS e-Infrastructure, supported in
part by FP7 projects EGI-InSPIRE, PRACE-1IP, PRACE-2IP, and HP-SEE.
The authors also acknowledge support received through SCOPES grant
IZ73Z0--128169 of the Swiss National Science Foundation.
\end{acknowledgments}



\begin{references}
\bibitem{my_2} The zeroth-order exponent for $\Delta^2_{L,r}$ can be calculated using successive slopes for all $r$
because condition $|b_0(r)| \gg |b_1(r)| L^{-\theta_1+\theta_0}$ is always satisfied.  The results for the zeroth-order
terms are verified by interpolation through anti-symmetrizing data points $(\Delta^2_{L,r}-\Delta^2_{L,1/r})/2$.
\end{references}
\end{document}